\documentclass[openacc]{rsproca_new}

\usepackage{amsfonts, amsmath, bm, amssymb,graphicx}

\newcommand{\beq}{\begin{equation}}
\newcommand{\enq}{\end{equation}}

\newcommand{\upi}{{\pi}}

\newcommand\I{\mbox{i}}
\newcommand\D{\mbox{d}}

\newcommand{\C}{\mathbb C}
\newcommand{\e}{\eqref}

\newcommand \bfr{{\bf r}}

\newcommand{\p}{\partial}

\begin{document}

\title{Short branch cut approximation in $2$D Hydrodynamics with Free Surface
}

\author{
 A. I. Dyachenko$^{1,2}$,
  S. A. Dyachenko$^{3,4}$,
        P. M. Lushnikov$^{1,5}$, and V. E. Zakharov$^{1,2,6}$}

\address{$^{1}$Landau Institute For Theoretical Physics,
Russia \\
$^{2}$Center for Advanced Studies, Skoltech, Moscow, 143026, Russia \\
$^{3}$Department of Applied Mathematics, University of Washington,
USA  \\
$^{4}$Department of Mathematics, SUNY Buffalo, 
USA  \\
$^{5}$Department of Mathematics and Statistics, University of New Mexico,
USA \\
$^{6}$Department of Mathematics, University of Arizona, USA }

\subject{fluid dynamics, water waves, conformal mapping}


\corres{P. M. Lushnikov\\
\email{plushnik@math.unm.edu}}

\begin{abstract}
A potential motion of ideal incompressible fluid with a free surface and
infinite depth is considered in two-dimensional geometry. A time-dependent
conformal mapping of the lower complex half-plane of the auxiliary complex
variable $w$ into the area filled with fluid is performed with the real
line of $w$ mapped into the free fluid's surface. The fluid dynamics can
be fully characterized by the motion of the complex singularities in the
analytical continuation of both the conformal mapping and the complex
velocity. We consider the short branch cut approximation of the dynamics
with the small parameter being the ratio of the length of the branch cut
to the distance between its center and the real line of $w$. We found that
the fluid dynamics in that approximation  is reduced to the complex Hopf
equation for the complex velocity coupled with the complex transport
equation for the conformal mapping. These equations are fully integrable
by characteristics producing the infinite family of solutions, including
 moving square root branch points and poles. These solutions involve  practical initial conditions resulting in jets and overturning waves.
The solutions are compared
with the simulations of the fully nonlinear Eulerian dynamics  giving
excellent agreement even when the small parameter approaches about one.
 \end{abstract}

\maketitle


\section{Introduction and basic equations }
\label{sec:introduction}

 \begin{figure}
\includegraphics[width=0.9859\textwidth]{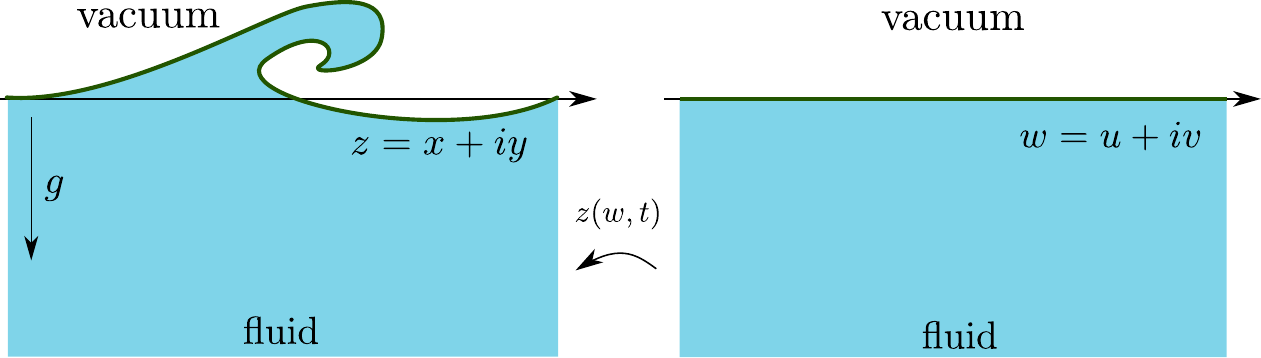}
\caption{ Shaded area represents the domain occupied by fluid both in the
physical plane $z=x+\I y $ (left) and  in the plane   $w=u+\I
v$  (right). Thick solid lines correspond to the fluid's free
surface. Gravity is  pointing downwards, i.e. in the direction $-y$ on the left panel.} \label{fig:schematic1}
\end{figure}

We consider a two-dimensional potential motion of
ideal incompressible fluid with the free surface of infinite depth in the gravity field $g$ as schematically shown on the left panel of
Fig. \ref{fig:schematic1}.
Fluid occupies the infinite region $-\infty < x < \infty$ in the
horizontal direction $x$ and extends down to $y\to -\infty$ in the
vertical direction $y$. We assume that there is no dependence on the third spatial dimension, i.e. the fluid motion is exactly two dimensional.     The bulk of fluid is at rest, i.e. there is no motion of fluid  both at
$|x|\to \pm\infty   $ and $y\to -\infty$.

We use a time-dependent conformal mapping
\begin{equation} \label{zwdef}
z(w,t)=x(w,t)+\I y(w,t)
\end{equation}
of the lower complex half-plane $\mathbb{C}^-$ of the auxiliary complex variable
\begin{equation} \label{wdef}
w\equiv u+\I v, \quad -\infty<u<\infty,
\end{equation}
into the area in $(x,y)$ plane occupied by the fluid. Here the real line $v=0$ is mapped into the fluid free surface (see Fig. \ref{fig:schematic1}) and $\mathbb{C}^-$ is defined by the condition  $-\infty<v\le0$.
 Then the time-dependent fluid free surface is represented in the parametric form as
\begin{equation} \label{xyu}
x=x(u,t), \ y=y(u,t).
\end{equation}
We assume a decay of perturbation of fluid beyond a flat surface $y\equiv0$ at  $x(u,t)\to \pm \infty$  which requires that %
\begin{equation} \label{zlimit}
z(w,t)\to w+o(1) \ \text{for} \ |w|\to\infty, \ w\in\C^-,
\end{equation}
where $o(1)$ means a vanishing contribution in that limit $|w|\to\infty$. The conformal mapping \e{zwdef} implies  that $z(w,t)$ is the analytic function of   $w\in\mathbb{C^-} $ and  %
\begin{equation} \label{zwconformal}
z_w\ne 0 \ \text{for any} \  w\in\mathbb{C^-},
\end{equation}
where  subscripts here and below means partial derivatives, $z_w\equiv\frac{\p z(w,t)}{\p w}$ etc.

Potential  fluid motion means
 that a velocity ${\bf v}$    of fluid is determined by a
velocity potential $\Phi(\bfr,t)$ as ${\bf v}= \nabla \Phi$ with $\nabla\equiv(\frac{\p}{\p x},\frac{\p}{\p y})$.  The
incompressibility condition $\nabla \cdot {\bf v} = 0$ results in
the
Laplace equation
\begin{align} \label{laplace}
\nabla^2 \Phi = 0
\end{align}
inside fluid, i.e. $\Phi$ is the harmonic function inside fluid.
 Eq. \e{laplace} is supplemented with  a decaying boundary condition (BC) at infinity,  %
\begin{equation} \label{phiinfinitya}
\nabla\Phi\to 0 \ \text{for } \ |x|\to \infty \ \text{or} \ y\to - \infty,
\end{equation}
which can be replaced without loss of generality by a zero  Dirichlet BC
\begin{equation} \label{phiinfinity}
\Phi\to 0 \ \text{for } \ |x|\to \infty \ \text{or} \ y\to - \infty
\end{equation}
The harmonic conjugate of $\Phi$ is a  stream  function $\Theta$ defined by
\begin{equation} \label{Thetadef}
\Theta_x=-\Phi_y \ \text{and} \ \Theta_y=\Phi_x.
\end{equation}
Similar to Eq. \e{phiinfinity}, we set without loss of generality a zero  Dirichlet BC for $\Theta$ as

\begin{equation} \label{Dirichlet2}
\Theta\to 0 \  \text{} \ \text{for } \ |x|\to \infty \ \text{or} \ y\to - \infty.
\end{equation}

 We define a complex velocity potential $\Pi(z,t)$~as%
\begin{equation} \label{ComplexPotentialdef}
\Pi=\Phi+\I \Theta,
\end{equation}
 where $z=x+\I y$ is the complex coordinate. Then Eqs. \e{Thetadef} turn into
Cauchy-Riemann equations resulting in the analyticity of $\Pi(z,t)$ in the domain of $z$ plane occupied by the fluid. A physical velocity with the components
$v_x$ and $v_y$ (in $x$ and $y$ directions, respectively) is obtained from $\Pi$ as $\frac{d\Pi}{dz}=v_x-\I v_y$.
The conformal mapping \e{zwdef} ensures that the function $\Pi(z,t)$~ %
\e{ComplexPotentialdef} transforms into $\Pi(w,t)$ which is the analytic function of $w$ for $w\in\mathbb{C^-}$ (in the bulk of
fluid).
Here and below we abuse the notation and use the  same symbols for functions of either  $w$ or $z$     (in other words, we  assume that e.g. $\tilde \Pi(w,t)= \Pi(z(w,t),t) $ and remove $\tilde ~$ sign).
The conformal transformation    \e{zwdef} also results in   Cauchy-Riemann equations $
\Theta_u=-\Phi_v, \quad \Theta_v=\Phi_u $
  in $w$ plane.

BCs at the free surface are time-dependent and consist of kinematic and dynamic BCs.
A kinematic BC states that  the free surface moves together with fluid particles located at that surface, i.e. there is no separation of fluid particles from the free surface. For mathematical formulation we look at the normal component of velocity $v_n$ (normal to the free surface) of such fluid particles.   Motion of the free surface is determined by a time derivative of the parameterization  \e{xyu}  so the kinematic BC is given by a projection into the normal direction as
\begin{equation} \label{kinematicu0}
{\bf n}\cdot\left(x_t,y_t \right )=v_n\equiv{\bf n}\cdot\nabla \Phi|_{x=x(u,t),\ y=y(u,t)},
\end{equation}
where %
${\bf n}=\frac{(-y_u,x_u)}{(x_u^2+y_u^2)^{1/2}}
$
is the outward unit normal vector to the free surface and subscripts here and below means partial derivatives, $x_t\equiv\frac{\p x(u,t)}{\p t}$ etc.

 A dynamic BC is given by the time-dependent Bernoulli equation (see e.g. Ref. \cite{LandauLifshitzHydrodynamics1989})
at the free surface,%
\begin{align} \label{dynamic1}
\left.\left(\Phi_t +
 \dfrac{1}{2}\left(\nabla \Phi\right)^2+gy\right)\right|_{x=x(u,t),\ y=y(u,t)}  = 0,
\end{align}
 where $g$ is the acceleration due to gravity. Here without loss of generality we assumed that pressure is zero above the free surface (i.e. in vacuum) which is ensured by the zero at the left-hand side (l.h.s.) of Eq. \e{dynamic1}.
All results below apply both to the surface gravity wave case ($g>0$) and the Rayleigh-Taylor problem $(g<0)$. We also consider a particular case $g=0$ when inertia forces  well exceed
 gravity force.

Eqs.
 \e{kinematicu0} and \e{dynamic1}  together with the analyticity (with respect to the independent variable $w$) of both $z(w,t)$ and $\Pi(w,t)$ inside fluid  as well as the decaying BCs \e{zlimit}, \e{phiinfinity} and \e{Dirichlet2}  form  a closed set of equations which is equivalent to Euler equations for dynamics of ideal fluid with free surface. The approach of using time-dependent conformal transformation like
\e{zwdef} to address free surface dynamics of ideal fluid    was exploited
by several authors
including~\cite{Ovsyannikov1973,MeisonOrzagIzraelyJCompPhys1981,TanveerProcRoySoc1991,TanveerProcRoySoc1993,DKSZ1996,ChalikovSheininAdvFluidMech1998,ChalikovSheininJCompPhys2005,ChalikovBook2016,ZakharovDyachenkoVasilievEuropJMechB2002}.
We follow the approach of Refs.
\cite{DKSZ1996,ZakharovDyachenkoArxiv2012,DyachenkoLushnikovZakharovJFM2019}
  to transform from the unknowns  $z(w,t)$ and $\Pi(w,t)$ into the equivalent  ``Tanveer-Dyachenko''  variables   %
\begin{align} \label{RVvar1}
&R=\frac{1}{z_w}, \\ \label{RVvar2}
&V=\I\frac{\p \Pi}{\p z}=\I R \Pi_w
\end{align}
These variables were introduced by S. Tanveer in Ref.
~\cite{TanveerProcRoySoc1991} for the periodic BC and later independently
obtained by A. I. Dyachenko in Ref.
\cite{Dyachenko2001} for the decaying  BCs  \e{zlimit}, \e{phiinfinity}
and \e{Dirichlet2} so we refer to these variables as ``Tanveer-Dyachenko
variables". Both
\begin{equation}\label{RVanalytic}
R(w,t) \ \text{and} \ V(w,t) \ \text{are the analytic functions for} \  w\in\C^-
\end{equation}
 as follows from Eq. \e{zwconformal} and the analyticity of both $z$\ and $\Pi$ for $w\in\C^-.$ Then the dynamical equations at the real line $w=u$ take the following complex form
 \cite{Dyachenko2001}: \begin{align}
\frac{\partial R}{\partial t} &= \I \left(U R_u - R U_u \right), \label{Reqn4}\\
\qquad U&=\hat P^-(R\bar V+\bar RV), \quad B= \hat P^-(V\bar V), \label{UBdef4}\\
\frac{\partial V}{\partial t} &= \I \left[ U V_u - RB_u\right ]+ g(R-1), \label{Veqn4}
\end{align}
where
\begin{equation} \label{Projectordef}
\hat P^-=\frac{1}{2}(1+\I \hat H)  \quad\text{and}\quad  \hat P^+=\frac{1}{2}(1-\I \hat H)
\end{equation}
 are the projector operators of any  function $q(u)$ (defined at the real
line $w=u$) into  functions $q^+(u)$ and $q^-(u)$ analytic in $w\in\mathbb{C}^-$ and
$w\in\mathbb{C}^+$, respectively, such that %
$q=q^++q^-,
$
%
i.e. $\hat P^+(q^++q^-)=q^+$ and   $\hat P^-(q^++q^-)=q^-.$
Here we assume that $q(u)\to 0$ for $u\to \pm\infty$.
Also the bar means complex conjugation and
\begin{equation} \label{Hilbertdef}
\hat H f(u)=\frac{1}{\upi} \text{p.v.}
\int^{+\infty}_{-\infty}\frac{f(u')}{u'-u}\D u'
\end{equation}
is the Hilbert
transform with $\text{p.v.}$ meaning a Cauchy principal value of the integral.
See also more discussion of the operators    \e{Projectordef}  in Ref.
\cite{DyachenkoLushnikovZakharovJFM2019}.
We refer to Eqs.  \e{Reqn4}-\e{Veqn4} as ``Dyachenko equations" for Tanveer-Dyachenko variables \e{RVvar1}-\e{RVvar2}. Note that Ref. ~\cite{TanveerProcRoySoc1991} also provided the dynamic equations for these variables for particular case of the periodic BC. These dynamic equations are written in terms of contour integrals with non-polynomial nonlinearities. Although in the periodic case these dynamics equations are equivalent to Dyachenko equations  \e{Reqn4}-\e{Veqn4}, we prefer to use the formulation  \e{Reqn4}-\e{Veqn4} because it has only cubic nonlinearity and avoid contour integrals by the projectors \e{Projectordef}. That formulation is valid both for the periodic BC and decaying BCs  \e{zlimit}, \e{phiinfinity} and \e{Dirichlet2}.

A complex conjugation $\bar f(w)$ of $f(w)$  in Eqs. \e{Reqn4}-\e{Veqn4} and throughout this paper is understood as applied with the assumption that $f(w)$ is the complex-valued function of the real argument $w$ even if $w$ takes the complex values so that %
\begin{equation} \label{bardef}
 \bar f(w)\equiv \overline {f(\bar{w})}.
\end{equation}That definition ensures
  the analytical continuation of $f(w)$ from
the real axis  $w=u$ into the complex plane of $w\in\mathbb{C.}$ Following Ref. \cite{DyachenkoDyachenkoLushnikovZakharovJFM2019},
we consider an analytical continuation of
the functions $R$\ and $V$ into the Riemann surfaces which we call by
$\Gamma_R(w)$ and  $\Gamma_V(w)$, respectively.

The decaying BCs  \e{zlimit}, \e{phiinfinity} and \e{Dirichlet2} imply that %
\begin{equation}\label{RVdecayingBC}
R(w,t)\to 1, \ V(w,t)\to 0 \ \text{for} \  \ |Re(w)|\to \infty \ \text{or} \ Im(w)\to - \infty.
\end{equation}
Also in Section \ref{sec:narrowbranchcutperiodic} we  consider the periodic BCs which are still decaying deep inside the fluid as
\begin{equation}\label{RVperiodicBC}
R(w+\lambda,t)= R(w,t),\, V(w+\lambda,t)= V(w,t), \, R(w,t)\to 1, \, V(w,t)\to 0 \ \text{for} \, Im(w)\to - \infty
\end{equation}
where the spatial period $\lambda$ can be set to $2\pi$\ without the loss of generality.

The variables  $R$ and $V$ \e{RVvar1} and \e{RVvar2} include only a derivative of the conformal mapping \e{zwdef} and the complex
potential $\Pi$ over $w$ while $z(w,t)$ and $\Pi(w,t)$ are recovered from solution of these Eqs. as $z=\int \frac{1}{R} dw$ and $\Pi=-\I\int \frac{V}{R}dw$.
It provides a freedom of adding arbitrary functions of time to both  $z(w,t)$ and $\Pi(w,t)$. Such addition is not physically important for $\Pi(w,t)$ because
it does not change fluid velocity while the addition to   $z(w,t)$  changes the location of the free surface. The freedom  in the real part $x$ is removed if one notices that generally for the decaying BCs \e{RVdecayingBC} $x(u,t)\to u+x_0$ for $|u|\to \infty $ and we choose the constant $x_0$ to be zero according to Eq. \e{zlimit}. The freedom  in the imaginary part $y$ is removed  from Eq. \e{zlimit}
by setting that  that  $y(u,t)\to 0$ for $|u|\to \infty$.

For the periodic BCs \e{RVperiodicBC}, we remove the freedom in $x$ by setting that%
\begin{equation}\label{xpi}
x(\pi,t)=\pi.
\end{equation}
  The freedom in the imaginary part $y$ is removed by ensuring the conservation of the total mass of the fluid. That conservation is expressed through the time independence of the following integral (see e.g. \cite{DyachenkoLushnikovZakharovJFM2019})
\begin{equation} \label{Mass2}
M=\int\limits^\pi_{-\pi}y(u,t) x_u(u,t)\D u.
\end{equation}
Without  loss of  generality, we set $M=0$, which corresponds to the zero mean level of the fluid in the vertical direction. E.g. the flat free surface would correspond  to  $y\equiv 0.$ Respectively, while recovering $z(w,t)$ from $R(w,t)$ for the periodic BCs we use that     $M=0$.

Ref. \cite{DyachenkoDyachenkoLushnikovZakharovJFM2019} found that the system
 \e{RVvar1}-\e{Veqn4}    has   an arbitrary number of pole solutions for $z_w$\ and $\Pi_w$. These poles are located for $w\in\C^+$, i.e. in the analytical continuation of   $z_w$\ and $\Pi_w$ to the area outside of the fluid domain.  These pole solutions allowed us to identify the existence of multiple nontrivial integrals of motion  (beyond the natural integrals like the Hamiltonian and the horizontal momentum), see Ref. \cite{DyachenkoLushnikovZakharovJFM2019} for details. Many of these integrals commute with respect to the non-canonical Poisson bracket found in Refs. \cite{ZakharovDyachenkoArxiv2012,DyachenkoLushnikovZakharovJFM2019}.
 It was suggested in  Ref. \cite{DyachenkoDyachenkoLushnikovZakharovJFM2019}  that the existence of such
commuting integrals of motion might be a sign of the Hamiltonian
integrability of the free surface hydrodynamics. It is well-established
(see e.g Refs.
\cite{DyachenkoLushnikovZakharovJFM2019,LushnikovZakharovWaterWaves2020,MooreProcRSocLond1979,MeironBakerOrszagJFM1982,BakerMeironOrszagJFM1982,KrasnyJFM1986,CaflischOrellanaSIAMJMA1989,CaflischOrellanaSiegelSIAMJAM1990,BakerShelleyJFM1990,ShelleyJFM1992,CaflishEtAlCPAM1993,BakerCaflischSiegelJFM1993,CowleyBakerTanveerJFM1999,BakerXieJFluidMech2011,Zubarev_Kuznetsov_JETP_2014,KarabutZhuravlevaJFM2014,ZubarevKarabutJETPLett2018eng,LushnikovZubarevPRL2018,LushnikovZubarevJETP2019})
that the system of the type
 \e{Reqn4}-\e{Veqn4} and its different generalizations also have solutions with branch points located for $w\in\C^+.$  Generally we expect coexistence of poles and branch points
 at different locations of   $w\in\C^+,$ see   Ref. \cite{DyachenkoDyachenkoLushnikovZakharovJFM2019} for numerical examples. Also Ref. \cite{LushnikovZakharovWaterWaves2020} demonstrated that purely rational solutions of the system
 \e{Reqn4}-\e{Veqn4}    are not very likely for the decaying boundary conditions \e{RVdecayingBC}. In particular, it was proven in Ref. \cite{LushnikovZakharovWaterWaves2020} that the rational solutions with the second  and/or first order poles  are impossible to survive with dynamics for any finite time duration, i.e. they are not persistent with time evolution.

The plethora of possible analytic solution makes it very important to find a tool to construct analytical solutions involving both branch points and poles. In this paper, we develop such a tool in the approximation of a short branch cut. Such approximations assume that the distance between the most remote branch points is much smaller than the distances from these points to the real line.  Then the
fluid dynamics is shown to be reduced to the complex Hopf equation for the
complex velocity coupled with the complex transport equation for the
conformal mapping. These equations are fully integrable by characteristics
producing the infinite family of solutions including the pairs of moving
square root branch points. We also provide an example of the excellent performance of the solution obtained in that approximation in comparison with  numerical solutions of the system
 \e{Reqn4}-\e{Veqn4} even when the length of the branch cut becomes comparable with its distance to the real line.


 The plan of the paper is the following.
 In Section \ref{sec:narrowbranchcut}  we derive the equations of the short branch cut approximation from the system
 \e{Reqn4}-\e{Veqn4}. The applicability condition of that approximation is also established. After that we show that these equations in the  moving complex frame are reduced to the fully integrable complex Hopf equation for the complex velocity $V(w,t)$ and the transport equation for $z(w,t)$.   Section \ref{sec:narrowbranchcutperiodic} develops  the short branch cut approximation for the spatially periodic case of  BCs \e{RVperiodicBC}.
Section \ref{sec:numericalcomparison} provides a comparison of the
analytical solutions of Section  \ref{sec:narrowbranchcutperiodic}  with
the full numerical solution of Eqs.  \e{Reqn4}-\e{Veqn4} and
\e{RVperiodicBC}. 
Section
\ref{sec:conclusion} gives a summary of obtained results and discussion of
future directions.

\section{ Short branch cut approximation and  square root singularity solutions}
\label{sec:narrowbranchcut}

In this section we derive the dynamical equations of the short  branch cut
approximation and establish their integrability in characteristics in subsection \ref{sec:Shortbranchcutapproximation}
as well as provide
particular solutions in subsection \ref{sec:Particularsolutions}.

\subsection{Short branch cut approximation}
\label{sec:Shortbranchcutapproximation}

 Consider the branch cut $\gamma$ connecting branch points at $w=a(t)\in\C^+$ and $w=b(t)\in\C^+$. The branch cut  is called short one if its distance
to the real axis, $\min(|Im(a)|,|Im(b)|)$, is large compared with  $|a-b|$. It allows to define  a small parameter $\epsilon$ as follows%
\begin{equation} \label{narrowbranchcut1}
\epsilon\equiv |a-b|/\min(|Im(a)|,|Im(b)|)\ll1.
\end{equation}

We neglect other singularities/branch cuts in $R$ and $V$ by assuming that they either identically zero or give small contribution at the real axis $w=u$. Then we define%
\begin{equation}\label{Rint}
\begin{split}
& R(w,t)-1=\int\limits^b_a \frac{\tilde R(w',t)\D w'}{w-w'}, \\
& V(w,t)=\int\limits^b_a \frac{\tilde V(w',t)\D w'}{w-w'},
\end{split}
\end{equation}
where $\tilde R(w',t)$ and  $\tilde V(w',t)$ are densities along branch
cut such that the jump of $R$  across branch cut at $w=w'$ is $2\pi \I
\tilde R(w',t)$ and similar the jump for $V$ is   $2\pi \I \tilde V(w',t)$
as follows from the Sokhotskii-Plemelj theorem  (see e.g.
\cite{Gakhov1966,PolyaninManzhirov2008}).
Integration in Eqs. \e{Rint} is taken over any contour which is a simple
arc in $\C^+$ connecting $w=a$ and $w=b.$ This contour  defines a branch cut. There is a freedom in choice of that branch cut connecting two branch points  $w=a$ and $w=b.$  We however assume that  the arclength of the branch  cut is of the same order of magnitude as $|a-b|$, i.e. that arclength is not very much different from the length of the segment of  the straight line connecting $w=a$ and $w=b.$ Also
$\tilde R(w',t)$ and  $\tilde V(w',t) $ are assumed to be the continuous
functions of $w'.$ Also $\tilde R(w',t)$ and  $\tilde V(w',t) $ can be
zero at some parts of the contour.  The functions $\bar R$ and $\bar V$
are given by
\begin{equation}\label{Rintbar}
\begin{split}
& \bar R(w,t)-1=\int\limits^{\bar b}_{\bar a} \frac{\bar {\tilde R}(\bar w',t)\D \bar w'}{w-\bar w'}, \\
& \bar V(w,t)=\int\limits^{\bar b}_{\bar a} \frac{\bar{\tilde V}(\bar
w',t)\D \bar w'}{w-\bar w'},
\end{split}
\end{equation}
with the contour $\bar \gamma $ connecting  $w=\bar a$ and $w=\bar b$
being the reflection of the contour of Eq.  \e{Rint} with respect to the
real axis $w=Re(w).$

Functions $U(w,t)$ and $B(w,t)$ can be rewritten as
\begin{equation}\label{UBPplus}
\begin{split}
& U=R\bar V+\bar RV-\hat P^+(R\bar V+\bar RV), \\
& B=V\bar V-\hat P^+(V\bar V),
\end{split}
\end{equation}
where we used the definition \e{Projectordef} to represent $\hat P^-$ as $\hat P^-=1-\hat P^+$.   Because  $\hat P^+ f$ is analytic for $w\in\C^+$  for any function $f$, as well as both $\bar R$ and $\bar V$  are analytic for $w\in\C^+$  according to the definition \e{bardef}, we conclude from Eq. ~\e{UBPplus} that both $U$\ and $B$\  have a branch cut $\gamma$
connecting $w=a$ and $w=b$   inherited from branch cut of $R$\ and $V.$
Then similar to Eqs. \e{Rint}, we represent $U(w,t)$ and $B(w,t)$
through the integrals of the densities $\tilde U(w',t)$ and  $\tilde
B(w',t)$ along the branch cut as
\begin{equation}\label{UBint}
\begin{split}
& U(w,t)=\int\limits^b_a \frac{\tilde U(w',t)\D w'}{w-w'}, \\
& B(w,t)=\int\limits^b_a \frac{\tilde B(w',t)\D w'}{w-w'}.
\end{split}
\end{equation}

 Using Eqs.  \e{Rint} and  \e{Rintbar}, a calculation of the projectors in the definitions \e{UBdef4} is performed through the partial fractions as follows%
\begin{align} \label{RVprojector}
&\hat P^- \left [  (R-1)\bar V\right ]=\hat P^-\int\limits^b_a \int\limits^{\bar b}_{\bar a} \frac{\tilde R(w'',t)\bar{\tilde V}(\bar w',t)\D w''\D \bar w'}{(w-w'')(w-\bar w')}\nonumber \\
&=\hat P^-\int\limits^b_a \int\limits^{\bar b}_{\bar a} \frac{\tilde
R(w'',t)\bar{\tilde V}(\bar w',t)\D w''\D \bar w'}{w''-\bar w'}\left
(\frac{1}{w-w''}-\frac{1}{w-\bar w'} \right )\nonumber\\
&=\int\limits^b_a \int\limits^{\bar b}_{\bar a} \frac{\tilde
R(w'',t)\bar{\tilde V}(\bar w',t)\D w''\D \bar w'}{w''-\bar
w'}\frac{1}{w-w''}=\int\limits^b_a  \frac{\tilde R(w'',t)\bar{ V}(w'',t)\D
w''}{w-w''},
\end{align}
where at the last line we used the definition \e{Rintbar}.  Similar to
Eq. \e{RVprojector}, one obtains  that \begin{align} \label{RbarVprojector}
&\hat P^- \left [  (\bar R-1) V\right ]=\int\limits^b_a  \frac{\tilde
V(w'',t)[\bar{ R}(w'',t)-1]\D w''}{w-w''}
\end{align}
and%
 \begin{align} \label{VVprojector}
& B(w,t)=\hat P^- \left [  V\bar V\right ]=\int\limits^b_a
\frac{\tilde V(w'',t)\bar{ V}(w'',t)\D w''}{w-w''}.
\end{align}

Eqs. \e{UBdef4},\e{UBint}-\e{VVprojector} result in %
\begin{equation} \label{Utilde1}
\tilde U(w,t)=\tilde V(w,t)\bar R(w,t)+\tilde R(w,t)\bar V(w,t)
\end{equation}
and \begin{equation} \label{Btilde1} \tilde B(w,t)=\tilde V(w,t)\bar
V(w,t),
\end{equation}
where $w\in \gamma.$

The functions $\bar R(w,t)$ and $\bar V(w,t)$ are analytic for $w\notin \bar \gamma $ including $w\in\C^+$ and they are represented by the convergent Taylor series in the open disk  $|w-w_0|<r_d$ with  $w_0\in \gamma.$ The radius of  convergence $r_d$   is given by distance from $w_0$ to $\bar \gamma$. For the short branch cut $r_d\simeq 2|a|\gg |b-a|$. Without the loss of generality we assume that the center of branch cut is located at the imaginary axis, i.e. $Re(a+b)=0$ and choose $w_0\in \gamma$ to be also at the imaginary axis, $Re(w_0)=0. $ E.g. for the simplest choice of branch cut $\gamma$ to be the segment of straight line connecting $w=a$ and $w=b,$ we  then obtain that%
\begin{equation}\label{w0def}
w_0=(a+b)/2.
\end{equation}
 In the short branch cut approximation \e{narrowbranchcut1} we keep only zeroth order terms in Taylor series for  $\bar R(w,t)$ and $\bar V(w,t)$ and denote %
\begin{equation} \label{RcVcdef}
R_c(t)\equiv \bar R(w_0(t),t) \ \text{ and} \ V_c(t)\equiv \bar
V(w_0(t),t).
\end{equation}
Using Eqs. \e{UBint},\e{Utilde1}-\e{RcVcdef} we then obtain in that
approximation that
\begin{equation}\label{UBnarrow}
\begin{split}
&U=RV_c+R_cV-V_c, \quad B=V_cV.
\end{split}
\end{equation}

More accurate approximation for $U$ and $B$ can be obtained from Eqs. \e{UBint},\e{Utilde1}-\e{RcVcdef}  by taking into account more terms in Taylor series of   $\bar R(w,t)$ and $\bar  V(w,t)$ at $w=w_0$ beyond Eq. \e{RcVcdef}. For instance, by keeping linear terms,%
\begin{equation}\label{VcRc1}
\begin{split}
& \bar R(w,t)\simeq R_c(t)+(w-w_0(t))R_c', \quad R_c'\left .\equiv \frac{\p}{\p w}R(w,t)\right |_{w=w_0}, \\
& \bar V(w,t)\simeq V_c(t)+(w-w_0(t))V_c', \quad V_c'\left .\equiv \frac{\p}{\p w}V(w,t)\right |_{w=w_0}, \\
\end{split}
\end{equation}
we obtain a modification of Eq.  \e{UBnarrow} as
$U\to U+\Delta U$ and $B\to B+\Delta B$, where%
\begin{equation}\label{Umodification}
\begin{split}
& \Delta U=-\langle \tilde R \rangle V_c'+ \langle \tilde V\rangle R_c'+(w-w_0)[V(w,t)R_c'+(R(\omega,t)-1)V_c']. \\
& \Delta B=-\langle \tilde V \rangle V_c'+(w-w_0)V(w,t)V_c'.
\end{split}
\end{equation}
Here $\langle \tilde R \rangle \equiv \int^b_a\tilde R(w)\D w$ and
$\langle \tilde V \rangle \equiv \int^b_a\tilde V(w)\D w$. The short
branch cut approximation requires that both
\begin{equation} \label{deltaUBcondition}
 |\Delta U|\ll |U| \quad \text{and} \quad |\Delta B|\ll |B|.
\end{equation}
 Qualitatively it implies that singularities in $R$ and $V$ must not be  too strong. E.g., if a singularity in $R$ is stronger than in $V$, as studied in \cite{DyachenkoDyachenkoLushnikovZakharovJFM2019}, then these conditions require that $|Im (a)V_c'\tilde R| \ll |R_c\tilde V|$. We note that the limit of infinitely short branch cut recovers pole solutions of \cite{DyachenkoDyachenkoLushnikovZakharovJFM2019}.

Any approximation of  $\bar R(w,t)$ and $\bar V(w,t)$ in Eqs.
\e{Utilde1},\e{Btilde1} by polynomials in powers of $w-w_0$ turns
Dyachenko Eqs. \e{Reqn4}-\e{Veqn4} into hyperbolic-type PDE with variable
coefficients both in $t$ and $w$. In the simplest case of zeroth order
polynomials, Eqs. \e{Reqn4}-\e{Veqn4}, \e{UBnarrow}
 and conditions \e{deltaUBcondition} result in the dynamical equations of the short branch cut approximation,%
\begin{equation}\label{RVnarrowdyn}
\begin{split}
& R_t+\I V_c R_u=\I R_c(V R_u-V_u R), \\
& V_t+\I V_c V_u=\I R_c VV_u+g(R-1),
\end{split}
\end{equation}
which have variable coefficients $R_c(t)$ and $V_c(t)$ in $t$ only. A more
general case of the higher order polynomials, i.e. going beyond the short
branch cut approximation implying variable coefficients in $w$ (as
exemplified in Eqs. \e{VcRc1}), will be considered in the separate paper.

In the complex moving frame,%
\begin{equation} \label{chidef}
\chi=w-\I\int\limits^t_0V_c(t')\D t',
\end{equation}
we obtain from Eqs. \e{RVnarrowdyn}, that
\begin{equation}\label{RVnarrowdynmoving}
\begin{split}
& R_t=\I R_c(V R_\chi-V_\chi R), \\
& V_t=\I R_c VV_\chi+g(R-1),
\end{split}
\end{equation}
where the space derivative is  over a new independent variable $\chi.$

We now neglect the term with $g$ in Eq. \e{RVnarrowdynmoving}, resulting
in
\begin{align}\label{Rnarrowdynmoving2}
& R_t=\I R_c(V R_\chi-V_\chi R), \\
& V_t=\I R_c VV_\chi, \label{Vnarrowdynmoving2}
\end{align}
 which is justified either if $g=0$ or $|R-1|\ll 1.$ This second condition implies that  the free surface is initially nearly flat  (this approximation applies only for small enough time while  the condition $|R-1|\ll 1$ remains valid).

Eq. \e{Vnarrowdynmoving2} is decoupled from Eq. \e{Rnarrowdynmoving2} and turns into the complex Hopf equation%
\begin{equation} \label{Hopfcompl}
V_\tau=V V_\chi
\end{equation}
under the transformation to the new complex time %
\begin{equation} \label{taudef}
\tau(t)=\I\int\limits^t_0R_c(t')\D t'
\end{equation}
and the respectively redefined Eq. \e{chidef} as
\begin{equation} \label{chidef2}
\chi=w-\int\limits^\tau_0\frac{V_c(t(\tau'))}{R_c(t(\tau'))}\D \tau'.
\end{equation}
Under the same transformation \e{taudef} and \e{chidef2}, Eq. \e{Rnarrowdynmoving2} turns into
\begin{equation} \label{Rnarrowdynmoving3}
R_\tau=VR_\chi-V_\chi R,
\end{equation}
which is convenient to transform back from $R$ \e{RVvar1} to $z$ which gives %
\begin{equation} \label{ztauchigeneral}
z_\tau=Vz_\chi+c(\tau).
\end{equation}
Eq. \e{ztauchigeneral} ensures that Eq. \e{Rnarrowdynmoving3} is valid for the arbitrary function  $c(\tau)$  of $\tau.$ To fix that freedom in  the choice of $c(\tau),$ we have, similar to the discussion after Eq. \e{RVperiodicBC} in Section \ref{sec:introduction},  to take into account the decaying BCs  \e{zlimit}  and \e{RVdecayingBC}. Using the definitions \e{taudef} and \e{chidef2}, we obtain that a change of independent variables from $(\chi,\tau)$ to $(w,t)$ in Eq. \e{ztauchi} results in
\begin{equation} \label{ztauchitw}
\frac{z_t}{\I R_c}+\frac{V_c}{R_c}z_w=Vz_w+c(\tau(t)).
\end{equation}
Taking the limit $w=u$, $u\to\pm \infty$, one obtains from  Eq.   \e{ztauchitw} and BCs \e{zlimit}, \e{RVdecayingBC} that %
\begin{equation}\label{ctaudecauing}
c(\tau)=\frac{V_c}{R_c},
\end{equation}
Respectively, Eq. \e{ztauchigeneral} is reduced to
\begin{equation} \label{ztauchi}
z_\tau=Vz_\chi+\frac{V_c}{R_c}.
\end{equation}

Eqs. \e{Hopfcompl} and \e{ztauchi} are easily integrable. Assume that $F(w)$ and $G(w)$ are arbitrary functions analytic for  $w\in \C^-$ such that $F(w)\to 0$ as $w\to\infty$ and $G(w)\to w$ as $w\to \infty$. Then a general solution of system \e{Hopfcompl} and \e{ztauchi} is given by %
\begin{align} \label{FGsol}
&V=F(\chi_0),\\
&z=G(\chi_0)+\int\limits^\tau_0c(\tau')\D\tau', \label{FGsol2}
\end{align}
where  the function $\chi_0(\chi,\tau)$ is determined by the solution of
the implicit equation
\begin{equation} \label{chi0eq}
\chi=\chi_0-F(\chi_0)\tau
\end{equation}
and
\begin{equation}\label{ctauint}
\int\limits^\tau_0c(\tau')\D\tau'=\int\limits^\tau_0\frac{V_c(\tau')}{R_c(\tau')}\D\tau'
\end{equation}
as follows from Eq. \e{ctaudecauing}.

Eqs. \e{FGsol} and \e{chi0eq} define a parametric representation of a
Riemann surface  $\Gamma_V(w)$. If $F(\chi_0)$ is the rational function
then   $\Gamma_V(w)$ has  genus zero at the initial time $t=0$ (see e.g.
Ref.  \cite{DubrovinFomenkoNovikovBookPartII1985} for definition of genus
of surface). For $t>0$, branch points emerge in  $\Gamma_V(w)$ thus making
genus nonzero. Branch points on the surface $\Gamma_V$ are zeros of the
derivative $\frac{d \chi}{d \chi_0}=1-F'(\chi_0)\tau$. Generally, these
zeros are simple. Assume such zero to be located at $\chi_0=\chi_c. $ Then
one can write that $\chi=(\chi_0-\chi_c)^2h(\chi_0)$ implying a square
root branch point on $\Gamma_V$ (one can solve that implicit equation for
$\chi_0(\chi)$ to see that).   Here $h(\chi_0)$ is the analytic function
of $\chi_0$  at $\chi_0=\chi_c $ such that $h(\chi_c)\ne 0.$  A number of
such branch points (and, respectively, the number of sheets of
$\Gamma_V(w))$ can be arbitrary large depending on the rational function
$F(\chi_0).$

A pair of Eqs. \e{FGsol} and \e{FGsol2}  give a parametric representation
of the ``physical" Riemann surface $G(z)$. This surface is not changing
with time meaning that a velocity field of the fantom fluid defined in
Ref. \cite{DyachenkoDyachenkoLushnikovZakharovJFM2019} is time-independent.  This fact
additionally shows that the short branch cut approximation has only a limited range of
applicability.

\subsection{Particular solutions}
\label{sec:Particularsolutions}

According to Refs.
\cite{DyachenkoDyachenkoLushnikovZakharovJFM2019,LushnikovZakharovWaterWaves2020},
 Eq. \e{FGsol}  does not allow decaying at $w\to \infty$ solution in terms of rational functions for $t>0$ because any $N$th order pole in $V$ immediately results in the $2N+1$  order pole term in
 the right-hand side (r.h.s.) of Eq. \e{FGsol} which cannot be balanced by the maximum $N+1$ pole order term in l.h.s. of  Eq \e{FGsol}. Assume that %
\begin{equation} \label{Feqini}
F(w)=-\frac{A}{w- a_0}=V|_{\tau=0},
\end{equation}
where $A$ and $a_0$ are the complex constants such that $a_0\in \C^+$. This
initial condition has a pole at $w=a_0$.
Then solving Eqs. \e{chi0eq} and \e{Feqini} for $\chi_0,$ we obtain that %
\begin{equation} \label{chi0tau}
\chi_0=\frac{\chi+a_0}{2}\pm\sqrt{\frac{(\chi-a_0)^2}{4}-A\tau}
\end{equation}
which has two square root branch points at %
\begin{equation} \label{chi012}
\chi=a_0\pm\sqrt{4A\tau}.
\end{equation}
We choose a branch cut to be the straight line segment of length
$|2\sqrt{4A\tau}|$ connecting two branch points \e{chi012}.

Eqs. \e{taudef},\e{chidef2},\e{FGsol},\e{chi0eq}-\e{chi0tau} result in%
\begin{equation} \label{chi-chi}
(\chi_0)_{\chi}=\frac{1}{2}+\frac{\chi-a_0}{4\sqrt{\frac{(\chi-a_0)^2}{4}-A\tau}}
\end{equation}
and
\begin{equation} \label{Vtauchi}
V=\frac{-2A}{\chi-a_0+\sqrt{(\chi-a_0)^2-4A\tau}}=-\frac{\chi-a_0-\sqrt{(\chi-a_0)^2-4A\tau}}{2\tau},
\end{equation}
where the branch of the square root  $\sqrt{\ldots}$ is chosen to have
$\sqrt{(\chi-a_0)^2}=\chi-a_0$ thus satisfying the initial condition
\e{Feqini}.

The length of the branch cut according to \e{chi012} is increasing with time as $2\sqrt{4A\tau}$ and the solution \e{Vtauchi} remains valid while the short cut approximation \e{narrowbranchcut1}  is valid, i.e.%
\begin{equation} \label{taucondition0}
|2\sqrt{4A\tau}| \ll |Im(a_0)|.
\end{equation}
That condition can be generalized by taking into account Eqs. \e{taudef}
and \e{chidef2}.

Eq. \e{FGsol2} for $z$ depends on the arbitrary function $G(\chi_0)$ so we can immediately construct the infinite set of solutions for $z$. E.g., choosing %
\begin{equation} \label{Gtrivial}
G(\xi)=\xi \ \text{for any} \ \xi\in \C,
\end{equation}
we obtain from Eq. \e{FGsol2} and \e{chi0tau} that %
\begin{equation} \label{zGtrivial}
z=\frac{\chi+a_0}{2}+\sqrt{\frac{(\chi-a_0)^2}{4}-A\tau}+ \int\limits^\tau_0\frac{V_c(\tau')}{R_c(\tau')}\D\tau'
\end{equation}
with the same choice of the branch of square root as in Eq. \e{Vtauchi}.
 Below in this section we always assume the same choice of the root. Using
the definition \e{RVvar1} we obtain from Eq. \e{chi0tau} that
\begin{equation} \label{Rfromz}
R=\frac{1}{(\chi_0)_{\chi} G_{\chi_0}(\chi_0)_{}}.
\end{equation}
Eqs. \e{Gtrivial}-\e{Rfromz}
result in%
\begin{equation} \label{Rztrivial}
R=\frac{2\sqrt{(\chi-a_0)^2-4A\tau}}{\chi-a_0+\sqrt{(\chi-a_0)^2-4A\tau}}=\frac{\sqrt{(\chi-a_0)^2-4A\tau}\left
(\chi-a_0-\sqrt{(\chi-a_0)^2-4A\tau}\right )}{2A\tau}.
 \end{equation}
This case corresponds to $R|_{t=0}\equiv 1$, which is the initially flat
free surface evolving from the initial velocity distribution \e{Feqini}. A
solution with such initial condition was first studied in Ref.
\cite{KuznetsovSpektorZakharovPhysLett1993,KuznetsovSpektorZakharovPRE1994}
in the approximation of weak
nonlinearity. It follows from Eq. \e{chi012} that one of two branch points reaches the real line $w=Re(w)$ in a finite time  for a general complex value of the complex constant$A$ (the only exception is $A>0$ when both branch cuts move horizontally parallel to the real line). It means a formation of singularity on the free surface. However, well before that the condition \e{narrowbranchcut1} of the applicability of the short branch cut approximation is violated as the lower branch point approaches the real line. In Section \ref{sec:numericalcomparison} we discuss such type of solution in details for the periodic boundary conditions and compare it with the full numerical solution of Euler equations indicating that the singularity in full equations does reach the real line in a finite time.

We now convert the solution \e{chi012} for the location of branch points into $w$ plane and the physical time $t.$ The location of  $w_0(t)$
\e{w0def}
is determined
by taking the midpoint %
\begin{equation}\label{chimiddef0}
\chi_{mid}\equiv a_0
\end{equation}
between the two branch points \e{chi012} and after that using the definitions \e{taudef} and \e{chidef2} to shift $\chi$ by $ \int\limits^\tau_0\frac{V_c(t(\tau'))}{R_c(t(\tau'))}\D \tau'$ to return from the independent variable  $\chi$  to $w.$ It gives that%
\begin{equation}\label{w0t}
w_0(t)=a_0+  \int\limits^\tau_0\frac{V_c(t(\tau'))}{R_c(t(\tau'))}\D \tau'.
\end{equation}

For $z$ we use the initial condition \e{Gtrivial} so that

\begin{equation}\label{Rini00}
R(w,t)|_{t=0}\equiv 1.
\end{equation}

In the simplest approximation of Eqs. \e{taudef} and \e{chidef2}, we set
\begin{equation} \label{chiw10}
\chi\simeq  w-\I V_c(0)t=w+ \I\frac{\bar At}{a_0-\bar a_0}
\end{equation}
and
\begin{equation}\label{taut10}
\tau\simeq \I R_c(0)t=\I \,t.
\end{equation}
where we used Eqs.  \e{taudef},\e{Feqini},\e{chi012},\e{w0t} and \e{Rini00}.

Using Eqs. \e{RcVcdef},\e{chi012},\e{chiw10} and  \e{taut10}, we obtain the approximate positions of branch points in $w$ as follows %
\begin{align} \label{chi012w}
&w=w_\pm\simeq a_0\pm \sqrt{ 4A\I \,t}- \I\frac{\bar At}{a_0-\bar a_0}.
\end{align}
It is shown in Section     \ref{sec:numericalcomparison}   that the periodic boundary conditions version of  equation \e{chi012w} is accurate for the values of $t$  well below the applicability condition \e{taucondition0}
of the short branch cut approximation. Thus it might be sufficient in many practical calculations  to use  Eq. \e{chi012w} instead of more accurate evaluations of integrals in Eqs. \e{taudef} and \e{chidef2}.

As another particular initial shape of surface we  choose that %
\begin{equation} \label{Glog}
G(\xi)=\xi+B\log[\xi-C] \ \text{for any} \ \xi\in \C \ \text{with} \ \quad
C\ne a_0 \ \text{and} \ Im(C)>0,
\end{equation}
where $B$ and $C$ are complex constants. We note that  Eq.  \e{Glog} does not satisfy BC \e{zlimit}. That asymptotic deficiency can be fixed if we add the extra term $-B\log[\xi-C_1], \ Im(C_1)>0$ in r.h.s. of   Eq.  \e{Glog} which is however beyond the scope of that paper. Respective, by ignoring such fix we also neglect the last term in r.h.s. of Eq. \e{FGsol2}. Then
 Eqs. \e{FGsol2} and \e{Glog}
imply that at any moment of time $\tau$,
\begin{equation} \label{zlog}
z=\chi_0(\chi,\tau)+B\log[\chi_0(\chi,\tau)-C], \
\end{equation}
where $\chi_0(\chi,\tau)$ is given by Eq. \e{chi0tau}.

Respectively, using Eq. \e{Rfromz}, we obtain from Eq. \e{zlog} that %
\begin{equation} \label{Rlog}
R=\frac{1}{(\chi_0)_{\chi}}\frac{\chi_0-C}{\chi_0-C+B}=\frac{1}{(\chi_0)_{\chi}}\left
(1-\frac{B}{\chi_0-C+B}\right )
\end{equation}
and
\begin{equation} \label{zwlog}
z_\chi={(\chi_0})_{\chi}\left (1+\frac{B}{\chi_0-C}\right ).
\end{equation}

We note that Eq. \e{Glog} has the branch cut which extends to the complex infinity.  However,  the corresponding $R$ at $t=0$ has only the pose singularity, see Eq.\e{Rlog}. Thus the short branch cut approximation remains valid for the initial condition \e{Glog}  at least for small enough $t$.

If $C\ne a_0$ then it follows from  Eq. \e{Glog} that the function $z_\chi$ has a simple pole at $\chi_0=C$.
Using Eq. \e{chi0tau}, we then obtain that a trajectory of motion of that
pole in $\chi$ plane is given by

\begin{equation} \label{chitrajectory}
\chi=C-\frac{A\tau}{a_0-C}.
\end{equation}
It follows from Eq. \e{zwlog} that the residue of $z_\chi$ at that point
is the integral of motion in $\chi$ plane, which is  exactly equal to the
constant $B$.

In a similar way, the function $R$ in Eq. \e{Rlog} has a simple pole at
$\chi_0=C-B$ provided  $C-B\ne a_0$. A trajectory of motion of that pole
in $\chi$ plane is given by

\begin{equation} \label{chitrajectory}
\chi=C-B-\frac{A\tau}{a_0-C+B}.
\end{equation}
However,  the residue of that pole is not a constant of motion. We notice
$V$ is regular at   $\chi_0=C-B$  (because we assumed $C-B\ne a_0$) at
least for small enough time. Such local solution  (with the pole in $R$
and no pole in $V$) is compatible with the analysis of
\cite{DyachenkoDyachenkoLushnikovZakharovJFM2019,LushnikovZakharovWaterWaves2020}
of the system  \e{Reqn4}-\e{Veqn4}, where solutions with the pole in both
$R$\ and  $V$ was excluded while a solution with the pole in $R$ only was
allowed.

Another particular case is to set  %
\begin{equation} \label{ReqiniTanveer}
R(w,t=0)=\frac{B_1}{w- a_0}=R|_{\tau=0},
\end{equation}
where $B_1$ is the complex constant. This initial condition has a pole at
the same $w=a_0$  as the initial pole in $V$  defined in Eq. \e{Feqini}.
Then Eqs.  \e{Gtrivial}-\e{Rfromz}
result in %
\begin{align}
&R=\frac{4B_1\sqrt{(\chi-a_0)^2-4A\tau}}{(\chi-a_0+\sqrt{(\chi-a_0)^2-4A\tau})^2}\nonumber \\
&=\frac{B_1\sqrt{(\chi-a_0)^2-4A\tau}\left
(\chi-a_0-\sqrt{(\chi-a_0)^2-4A\tau}\right
)^2}{4A^2\tau^2}.\label{RzTanveer}
\end{align}
The particular solution \e{Vtauchi},\e{RzTanveer} recovers the asymptotic
result of Ref. \cite{TanveerProcRoySoc1993} (Case (a) of Section 4 of that
Ref.) obtained in that Ref. by the matched asymptotic expansions at $t\to
0.$

Eq. \e{RzTanveer} makes sense locally near the pole position   but cannot
be valid globally because $R$ must approach 1 as $w\to \infty$. So we
provided that case only for the exact comparison with Ref.
\cite{TanveerProcRoySoc1993}. We however can easily fix that deficiency
through the replacement of Eq. \e{ReqiniTanveer}
by    %
\begin{equation} \label{Reqini}
R(w,t=0)=1+\frac{B_1}{w- a_0}=R|_{\tau=0}.
\end{equation}
Using Eq. \e{RVvar1} we then obtain that %
\begin{align} \label{ztanveer}
z(w,t=0)=w-B_1\log(w-a_0+B_1)
 \end{align}
and, using Eqs. \e{FGsol2},\e{chi0tau}, that
\begin{equation} \label{zR2}
z=\frac{\chi+a_0}{2}+\sqrt{\frac{(\chi-a_0)^2}{4}-A\tau}-B_1\log\left
(\frac{\chi-a_0}{2}+\sqrt{\frac{(\chi-a_0)^2}{4}-A\tau}+B_1 \right).
\end{equation}
Differentiating  Eq. \e{zR2}  over $w$ and using Eq. \e{RVvar1} results in
\begin{align}
&R=\frac{2\sqrt{(\chi-a_0)^2-4A\tau}(2B_1+\chi-a_0+\sqrt{(\chi-a_0)^2-4A\tau})}{(\chi-a_0+\sqrt{(\chi-a_0)^2-4A\tau})^2}\nonumber \\
&=\frac{\sqrt{(\chi-a_0)^2-4A\tau}(2B_1+\chi-a_0+\sqrt{(\chi-a_0)^2-4A\tau})\left
(\chi-a_0-\sqrt{(\chi-a_0)^2-4A\tau}\right
)^2}{8A^2\tau^2}.\label{RzTanveer1}
\end{align}
Eq. \e{RzTanveer} is recovered from Eq. \e{RzTanveer1} in the limit
$B_1\to \infty$.

We note that in all particular cases \e{Rztrivial}, \e{RzTanveer} and
\e{RzTanveer1}, the series expansion at any of two branch points
\e{chi012} shows that $R=0$ at these points, which is in the perfect
agreement with the analytical results of Ref.
\cite{LushnikovZakharovWaterWaves2020}. We also remind that all these
particular cases share the same $V$ from Eq. \e{Vtauchi}.

To express $V$ and $R$ in all these cases in terms of $w$ and $t$ requires
to find expression of $\chi$ and $\tau$ through $w$ and $t$ using Eqs.
\e{taudef} and \e{chidef2}. For that one can use definitions  \e{RcVcdef}
with $\chi_0$  determined in terms of $\chi$ through Eq. \e{chi012}. After
that a general condition \e{narrowbranchcut1} can be also verified. We note that all particular examples above correspond to the moving branch points according to Eq. \e{chi012}. It implies that the condition \e{narrowbranchcut1} is violated at large times so the short branch cut approximation is valid in all these particular cases only for a finite duration of time.

The second sheet of Riemann surface  $\Gamma_V$ corresponds to the opposite choice of
sign in Eq. \e{chi0tau}. It means that we have to change the sign in front of
each square root in Eqs. \e{Vtauchi},\e{Rztrivial},\e{RzTanveer} and  Eq.
\e{RzTanveer1}. It immediately implies that  $V\to \infty $ and $\
R\to\infty$ as $\chi\to \infty$ in all these equations for the second
(nonphysical) sheet of Riemann surface.

In all particular examples of this Section, the functions $V$ and $R$ are
analytic functions of $\sqrt{(\chi-a_0)^2-4A\tau}$, i.e. they are analytic
in two sheets of Riemann surface of $w$. This fact is the result of the
approximation \e{RVnarrowdyn} effectively assuming that both $\tilde V$
and $\tilde R$ are constant on the branch cut. Going beyond that short cut
approximation, we expect that $V$ and $R$ can be analytically continued
into a much more complicated Riemann surfaces $\Gamma_V(w)$ and
$\Gamma_R(w)$ with the unknown total number of sheets. Our experience with
the  Stokes wave in Ref. \cite{LushnikovStokesParIIJFM2016} suggests that
generally the number of sheets is infinite. Some exceptional cases like
found in Refs.
\cite{KarabutZhuravlevaJFM2014,ZubarevKarabutJETPLett2018eng} have only a
finite number of sheets of Riemann surface (these solutions however have
diverging values of $V$ and $R$ at $w\to\infty$). We suggest that the
detailed study of such many- and infinite-sheet Riemann surfaces is one of
the most important goal in free surface hydrodynamics. This topic is
however beyond the scope of this paper. We also notice that even in the
simplest considered case \e{Feqini}, the function $R$ can have the
arbitrary number of additional poles and branch points depending on the
choice of the function $G(\chi_0)$ in Eq. \e{FGsol} instead of particular
cases considered in this Section.

\section{ Short branch cut approximation and  square root singularity solutions for the periodic case}
\label{sec:narrowbranchcutperiodic}

In this section we extend the results of Section \ref{sec:narrowbranchcut}
into the  $2\pi$-periodic boundary conditions \e{RVperiodicBC} instead of the
decaying boundary conditions  \e{RVdecayingBC} used in Section \ref{sec:narrowbranchcut}.
 In that periodic case, instead of a single branch cut connecting branch points at $w=a(t)\in\C^+$ and $w=b(t)\in\C^+$  of Section \ref{sec:narrowbranchcut},
 we consider the periodic sum of branch cuts and use the identity %
\begin{align} \label{polessum}
\sum\limits^{\infty}_{n=-\infty}\frac{1}{n+a}=\upi\cot{\upi a.}
\end{align}
Then taking the sum over branch cuts amounts to replacing $w-w'$ by $(1/2)\tan{[(w-w')/2]}$ in the denominators of Eq. \e{Rint} and all other similar expressions.  In particular, Eq. \e{Rint} is replaced by%
\begin{equation}\label{Rintperiodic}
\begin{split}
& R(w,t)-1=\frac{1}{2}\int\limits^b_a \frac{\tilde R(w',t)\D w'}{\tan\frac{w-w'}{2}}, \\
& V(w,t)=\frac{1}{2}\int\limits^b_a \frac{\tilde V(w',t)\D w'}{\tan\frac{w-w'}{2}},
\end{split}
\end{equation}
 Eq. \e{Rintbar} is replaced by
\begin{equation}\label{Rintbarperiodic}
\begin{split}
& \bar R(w,t)-1=\frac{1}{2}\int\limits^{\bar b}_{\bar a} \frac{\bar {\tilde R}(\bar w',t)\D \bar w'}{\tan\frac{w-\bar w'}{2}}, \\
& \bar V(w,t)=\frac{1}{2}\int\limits^{\bar b}_{\bar a} \frac{\bar{\tilde
V}(\bar w',t)\D \bar w'}{\tan\frac{w-\bar w'}{2}}.
\end{split}
\end{equation}
and Eq. \e{UBint} is replaced by%
\begin{equation}\label{UBintperiodic}
\begin{split}
& U(w,t)=\frac{1}{2}\int\limits^b_a \frac{\tilde U(w',t)\D w'}{\tan\frac{w- w'}{2}}, \\
& B(w,t)=\frac{1}{2}\int\limits^b_a \frac{\tilde B(w',t)\D w'}{\tan\frac{w- w'}{2}}.
\end{split}
\end{equation}

 Instead of the partial fractions used in Eq. \e{RVprojector}, it is more convenient to use the integral representation  of the  projector $\hat P^-$ \e{Projectordef} for the periodic functions (see e.g.  Ref. \cite{LushnikovDyachenkoSilantyevProcRoySocA2017}) which follows from Eq. \e{Hilbertdef} and the  Sokhotskii-Plemelj theorem  (see e.g.
\cite{Gakhov1966,PolyaninManzhirov2008}) giving that %
\begin{align} \label{Puminusperiod}
\hat P^-f=-\frac{1}{2\pi \I}\sum \limits _{n=-\infty}^\infty \int\limits^{\pi}_{-\pi}\frac{f(u')du'}{u'-u+\I  0+2\pi n}=-\frac{1}{4\pi \I} \int\limits^{\pi}_{-\pi}\frac{f(u')du'}{\tan{\frac{u'-u+\I  0}{2}}},
\end{align}
where $\I 0$ means $\I \epsilon, \ \epsilon\to 0^+$ and we used the identity \e{polessum}.

Using Eqs.  \e{Rintperiodic},  \e{Rintbarperiodic} and \e{UBintperiodic}, a calculation of the projectors in the definitions \e{UBdef4} is performed through moving the integration contour from $(-\pi,\pi)$ to  $(-\pi-\I \infty,\pi-\I\infty)$ together with the identity $\tan(-\I \infty)=-\I$ which give
that %
\begin{align} \label{RVprojectora}
&\hat P^- \left [  (R-1)\bar V\right ]=\frac{1}{4}\hat P^-\int\limits^b_a \int\limits^{\bar b}_{\bar a} \frac{\tilde R(w'',t)\bar{\tilde V}(\bar w',t)\D w''\D \bar w'}{\tan\frac{w- w''}{2}\tan\frac{w-\bar w'}{2}}\nonumber \\
&=-\frac{1}{16\pi \I} \int\limits^{\pi}_{-\pi}\int\limits^b_a \int\limits^{\bar b}_{\bar a} \frac{\tilde
R(w'',t)\bar{\tilde V}(\bar w',t)\D w''\D \bar w' \D u'}{\tan\frac{u'- w''}{2}\tan\frac{u'-\bar w'}{2}\tan{\frac{u'-w+\I  0}{2}}}\nonumber\\
&=-\frac{1}{4 }\int\limits^b_a \int\limits^{\bar b}_{\bar a} \frac{\tilde
R(w'',t)\bar{\tilde V}(\bar w',t)\D w''\D \bar w'}{1}\left (\frac{-1}{2}-\frac{
1}{\tan\frac{w- w''}{2}\tan\frac{w-\bar w'}{2}}+\frac{
1}{\tan\frac{\bar w'- w''}{2}\tan{\frac{w-\bar w'}{2}}} \right )\nonumber\\
&=-\frac{1}{4 }\int\limits^b_a \int\limits^{\bar b}_{\bar a} \frac{\tilde
R(w'',t)\bar{\tilde V}(\bar w',t)\D w''\D \bar w'}{1}\left (\frac{-1}{2}-\frac{1}{\tan\frac{w-\bar w'}{2}}\left [\frac{
\tan\frac{\bar w'- w''}{2}-\tan\frac{w- w''}{2}}{\tan\frac{w- w''}{2}\tan\frac{\bar w'- w''}{2}}\right ] \right )\nonumber\\
&=-\frac{1}{4 }\int\limits^b_a \int\limits^{\bar b}_{\bar a} \frac{\tilde
R(w'',t)\bar{\tilde V}(\bar w',t)\D w''\D \bar w'}{1}\left (\frac{1}{2}+\left [\frac{
1}{\tan\frac{w- w''}{2}\tan\frac{\bar w'- w''}{2}}\right ] \right )\nonumber\\
&=-\frac{1}{8 }\int\limits^b_a \int\limits^{\bar b}_{\bar a} \frac{\tilde
R(w'',t)\bar{\tilde V}(\bar w',t)\D w''\D \bar w'}{1}+\frac{1}{4 }\int\limits^b_a \int\limits^{\bar b}_{\bar a} \frac{\tilde
R(w'',t)\bar{\tilde V}(\bar w',t)\D w''\D \bar w'}{\tan\frac{w- w''}{2}\tan\frac{w''-\bar w'}{2}},
\end{align}
where  we used the following  trigonometric identity
\begin{align} 
\cot{(a-b)}=\frac{1+\tan{a}\tan{b}}{\tan{a}-\tan{b}} \nonumber
\end{align} with $a=\frac{\bar w'- w''}{2},$  $b=\frac{w- w''}{2}$ and $a-b=\frac{\bar w'-w}{2}$.

Now using the definitions  \e{Rintperiodic}  and \e{Rintbarperiodic} in  Eq. \e{RVprojectora}, we obtain that

\begin{align}\label{RVprojectorperiodic}
&\hat P^- \left [  (R-1)\bar V\right ]=-\frac{1}{2 }[R(-\I \infty,t)-1]\bar V(\I \infty,t)+\frac{1}{2 }\int\limits^b_a \limits \frac{\tilde
R(w'',t)\bar{ V}( w'',t)\D w''}{\tan\frac{w- w''}{2}},
\end{align}
where at the first term in r.h.s. of Eq. \e{RVprojectora} we take appropriate limits to use the analyticity of $R$ and $\bar V$ as follows: $\tan\frac{w-w'}{2}\to -\I \infty$ as $w\to-\I\infty$
in the first  Eq.   \e{Rintperiodic}    $\tan\frac{w-\bar w'}{2}\to \I \infty$ as $w\to\I\infty$
in the second  Eq.   \e{Rintbarperiodic}.

 Similar
to  Eq. \e{RVprojectorperiodic}, one obtains  that \begin{align}
\label{RbarVprojectorperiodic} &\hat P^- \left [  (\bar R-1) V\right
]=-\frac{1}{2 }[\bar R(\I \infty,t)-1] V(-\I \infty,t)+\frac{1}{2 }\int\limits^b_a \limits \frac{\tilde V(w'',t)[\bar{ R}(w'',t)-1]\D w''}{\tan\frac{w- w''}{2}}
\end{align}
and%
 \begin{align} \label{VVprojectorperiodic}
&\hat B(w,t)=\hat P^- \left [  V\bar V\right ]=-\frac{1}{2 }[\bar V(\I \infty,t)] V(-\I \infty,t)+\frac{1}{2 }\int\limits^b_a
\frac{\tilde V(w'',t)\bar{ V}(w'',t)\D w''}{\tan\frac{w- w''}{2}}.
\end{align}

We obtain from Eqs. \e{bardef} and \e{RVperiodicBC} that

\begin{equation}\label{RVinfinity}
R(-\I \infty,t)-1 =\bar R(\I \infty,t)-1=\bar V(\I \infty,t)]=V(-\I \infty,t)=0.
\end{equation}
Then Eqs. \e{UBdef4},\e{UBintperiodic}-\e{VVprojector} and \e{RVinfinity} result in  the same Eqs. \e{Utilde1} and \e{Btilde1} as for the decaying BCs case \e{RVdecayingBC} considered in Section \ref{sec:narrowbranchcut}.

Similar to Section \ref{sec:narrowbranchcut}, we  consider the short branch cut approximation for the periodic case recovering  exactly the same Eqs.  as \e{RcVcdef}-\e{ztauchitw}.
The only difference in addressing these equations in comparison with  Section \ref{sec:narrowbranchcut} is to use the periodic BC \e{RVperiodicBC}. 
Respectively, instead of Eq. \e{ctaudecauing}, we have to use the conditions \e{xpi} and \e{Mass2} to determine  $c(\tau)$.

As a particular example, we assume a periodic initial condition %
\begin{equation} \label{Feqiniperiodic}
F(w)=-\frac{A}{2\tan{\frac{w- a_0}{2}}}+\frac{\I A}{2}=V|_{\tau=0},
\end{equation}
where $A$ and $a_0$ are complex constants such that $a_0\in \C^+$.
This initial condition is the periodic analog of  Eq. \e{Feqini} with the extra constant term $\frac{\I A}{2}$  added to make sure that $V\to 0$ at $Im(w)\to-\I\infty,$ i.e. the decay of the velocity deep inside fluid.   This initial condition has  poles at $w=a_0+2\pi n, \ n=0,\pm1,\pm2,\ldots$.
 In contrast with Eqs. \e{chi0eq} and \e{Feqini}, Eqs. \e{chi0eq} and \e{Feqiniperiodic} cannot be explicitly solved for  $\chi_0.$ Thus Eqs. \e{FGsol}-\e{chi0eq} provide only the implicit form of the solution for the initial condition \e{Feqiniperiodic}. Note that $c(\tau)$ in this section is generally not
given by Eq. \e{ctaudecauing} but is determined the conditions \e{xpi} and \e{Mass2}.

We can still explicitly
obtain that the locations of
square root branch points if we differentiate Eq. \e{chi0eq} over $\chi$ resulting in
\begin{equation}\label{Fder1}
1=(\chi_0)_{\chi}\left[ 1+\frac{dF(\chi_0)}{d\chi_0}\tau \right]
\end{equation}
and notice $(\chi_0)_{\chi}$ is singular at the square root branch points (see e.g. Eq. \e{chi-chi} in the non-periodic case). It implies from Eq. \e{Fder1} that
\begin{equation}\label{Fder2}
1+\frac{dF(\chi_0)}{d\chi_0}\tau=0
\end{equation}
 at each branch point. Solving Eq.  \e{Fder2} we obtain the location of branch point in $\chi_0$ variable as follows %
\begin{equation}\label{chi0branch1}
\chi_0=\chi_{0,\pm}\equiv a_0\pm 2\, \text{arcsin}\left(\frac{\sqrt{ A\tau}}{2 }\right) +2\pi n, \ n=0,\pm1,\pm2,\ldots.
\end{equation}

 Then using Eqs.  \e{Feqiniperiodic} ,\e{Fder2}   and  \e{chi0branch1}   we obtain that  the square root branch points  are located at %
\begin{equation} \label{chi012periodic}
\chi=\chi_{\pm}=a_0-\frac{\I A\tau}{2}\pm\left [ \frac{1}{2} \sqrt{ A\tau}  \sqrt{4- A\tau}+2\, \text{arcsin}\left(\frac{\sqrt{ A\tau}}{2 }\right) \right ]+2\pi n, \ n=0,\pm1,\pm2,\ldots.
\end{equation}
Eq. \e{chi012} is recovered from Eq. \e{chi012periodic} at the leading order $O(\tau^{1/2})$ for  $A\tau\to 0$ and $n=0.$ Similar to Eq.  \e{chi012}, we choose a branch cut to be the straight line segment of length
$| \sqrt{ A\tau}  \sqrt{4- A\tau}+4\, \text{arcsin}\left(\frac{\sqrt{ A\tau}}{2 }\right)|$ connecting the two branch points \e{chi012periodic}. The location of  $w_0(t)$
\e{w0def}
is determined
by taking the midpoint %
\begin{equation}\label{chimiddef}
\chi_{mid}\equiv a_0-\frac{\I A\tau}{2}
\end{equation}
between the two branch points \e{chi012periodic} and after that using the definitions \e{taudef} and \e{chidef2} to shift $\chi$ by $ \int\limits^\tau_0\frac{V_c(t(\tau'))}{R_c(t(\tau'))}\D \tau'$ to return from the independent variable  $\chi$  to $w.$ It gives that%
\begin{equation}\label{w0tperiodic}
w_0(t)=a_0-\frac{\I A\tau}{2}+  \int\limits^\tau_0\frac{V_c(t(\tau'))}{R_c(t(\tau'))}\D \tau'.
\end{equation}

For $z$ we use the initial condition \e{Gtrivial} so that

\begin{equation}\label{Rini0}
R(w,t)|_{t=0}\equiv 1.
\end{equation}

The length of the branch cut is increasing with time as $\propto \sqrt{\tau}$ at  $\sqrt{A\tau}$ according to  \e{chi012periodic}   and the solution \e{Vtauchi} remains valid at least while the short cut approximation \e{narrowbranchcut1}  is valid, i.e.%
\begin{equation} \label{taucondition}
|2\sqrt{4A\tau}| \ll |Im(a_0)|.
\end{equation}

For comparison with simulations one has to return from the independent variables  $\tau$ and $\chi$ to the original variables $t$ and $w$ using Eqs. \e{taudef} and \e{chidef2}.
In the simplest approximation of Eqs. \e{taudef} and \e{chidef2}, we set
\begin{equation} \label{chiw1}
\chi\simeq  w-\I V_c(0)t=w-\left. \I\left [-\frac{\bar A}{2\tan{\frac{w-\bar  a_0}{2}}}-\frac{\I\bar  A}{2}\right ]\right|_{w=a_0, t=0}\times t=w- \I\left [-\frac{\bar A}{2\tan{\frac{a_0-\bar  a_0}{2}}}-\frac{\I\bar  A}{2}\right ]t
\end{equation}
and
\begin{equation}\label{taut1}
\tau\simeq \I R_c(0)t=\I \,t.
\end{equation}
where we used Eqs.  \e{taudef},\e{chi012},\e{Feqiniperiodic},\e{w0tperiodic} and \e{Rini0}.

Using Eqs. \e{RcVcdef},\e{chi012periodic},\e{chiw1} and  \e{taut1}, we obtain the approximate positions of branch points in $w$ as follows %
\begin{align} \label{chi012periodicw}
&w=w_\pm\simeq a_0+ \I\left [-\frac{\bar A}{2\tan{\frac{a_0-\bar  a_0}{2}}}-\frac{\I\bar  A}{2}\right ]t+\frac{ At}{2}\nonumber \\ &\qquad\pm\left [ \frac{1}{2} \sqrt{ A\I \,t}  \sqrt{4- A\I \,t}+2\, \text{arcsin}\left(\frac{\sqrt{ A\I \,t}}{2 }\right) \right ]+2\pi n, \ n=0,\pm1,\pm2,\ldots.
\end{align}

For the precise location of branch points instead of the approximation \e{chi012periodicw}, we have to find the dependence of $\tau$ on $t$ and $\chi$ on $w$ and $t $ using Eqs. \e{taudef} and \e{chidef2} together with the definitions \e{RcVcdef} and Eq. \e{w0tperiodic}. These equations are implicit ones. Also one has to find $\chi_0(\chi,\tau) $ from the implicit Eqs. \e{chi0eq} and  \e{Feqiniperiodic} to be able to use Eqs. \e{FGsol} and \e{FGsol2} for finding     $V_c$ and $R_c$ from the definitions \e{RcVcdef}.

Similar to the solutions of Section \ref{sec:narrowbranchcut}, the particular solutions considered in this section also have the moving branch points according to Eq. \e{chi012periodic}.  It follows from   Eq. \e{chi012} that one of two branch points reaches the real line $w=Re(w)$ in a finite time  for a general complex value of the complex constant $A$. It means a formation of singularity on the free surface. However, well before that the condition \e{narrowbranchcut1} of the applicability of the short branch cut approximation is violated as the lower branch point approaches the real line.

Similar to the discussion in  Section \ref{sec:narrowbranchcut}, one can find a wide range of particular solutions for the periodic case of this section based on the general solutions \e{FGsol} and \e{FGsol2}.

\section{Comparison of short branch cut approximation with full numerical solution }
\label{sec:numericalcomparison}

In this section we compare the short branch cut approximation described in the
Section~\ref{sec:narrowbranchcutperiodic} with the full numerical solution of
the system~\e{Reqn4}--\e{Veqn4} satisfying the initial conditions~\e{Feqiniperiodic}
and~\e{Rini0}. We assume that there is no gravity ($g=0$). Both functions $z(w,t)$
and $\Pi(w,t)$ are recovered from the variables $R$ and $V$ by means of the
relations~\e{RVvar1} and~\e{RVvar2} as discussed in Section \ref{sec:introduction},
where we assume zero mean fluid~\e{Mass2} is at zero.

\begin{figure}
\begin{center}
\includegraphics[width=0.495\textwidth]{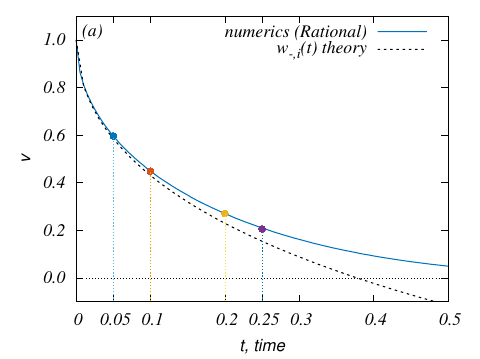}
\includegraphics[width=0.495\textwidth]{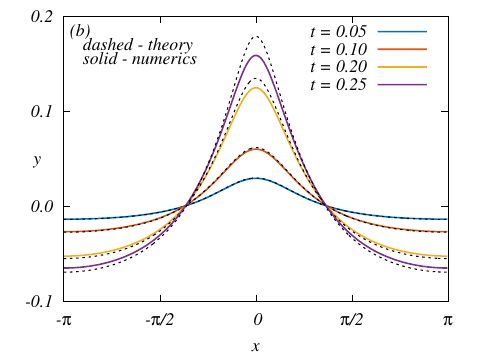}
\includegraphics[width=0.495\textwidth]{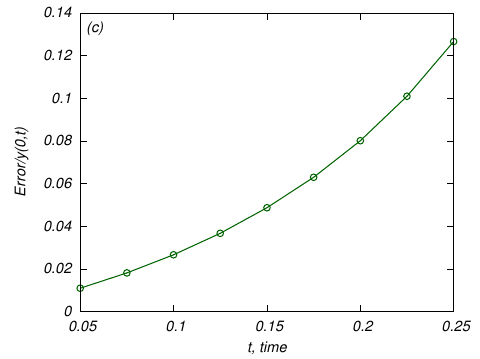}
\end{center}
\caption{(a) The vertical position $v=Im(w_{-}(t))=w_{-,i}(t)$ of the lower branch
point vs. time $t$ in the simulation with initial
conditions~\e{Feqiniperiodic} and \e{Rini0} with $A=\I$ and $a_0=\I$. The
lower branch point location $w_{-}(t)$ is recovered from the full numerical
simulations of Eqs.~\e{Reqn4}--\e{Veqn4} by means of a numerical rational
approximation (solid  line) compared with its location from the short branch cut
approximation~\e{chi012periodic} (dashed line). See Appendix  \ref{sec:Rationalapproximation}
about the rational approximation.  The relative error of the theoretic prediction vs. numerics is
$1.93\%$ at short time $t = 0.05$, about $3.93\%$ at $t = 0.1$, and
$15.2\%$ at $t= 0.2$.
At $t=0.05$ and $t=0.2$, the value of the parameter $\epsilon$
introduced in Eq.~\e{narrowbranchcut1} is $0.61$ and $0.89$ respectively.
These values are well outside the asymptotic condition $\epsilon\ll 1$ when
the short branch cut theory is guaranteed to be applicable.
(b) The spatial profiles of the fluid surface at different times: the result of
numerical simulation (solid lines), and the short branch cut approximation (dashed
lines). (c) The time dependence of the maximum of the error    for the surface elevation $y(x,t)$ between the numerical solution and the  short branch cut approximation. The maximum of error occurs at $x=0$ as seen from the surface profiles in (b).  The error is normalized to  the values of $y(0,x)$ from the numerical solution.
} \label{Ai1}
\end{figure}

\begin{figure}
\begin{center}
\includegraphics[width=0.495\textwidth]{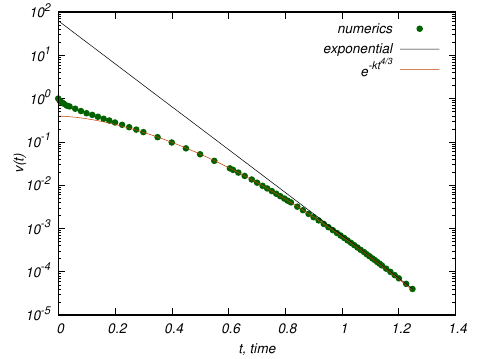}
\end{center}
\caption{A fit of  $v=w_{-,i}(t)$ from Fig. \ref{Ai1}a (shown by dots) into the stretched exponent $v=a e^{-k t^{b}}$ with $b=1$ (black solid line) and $b=4/3$ (red solid line). It is seen that $b=4/3$ is much better fit that $b=1$. Here $a= 0.39795\pm  0.01406$ and $k=6.40096 \pm 0.03348$ are the fitting constants.
}
\label{fig:stretchedexponent}
\end{figure}

\begin{figure}
\includegraphics[width=0.495\textwidth]{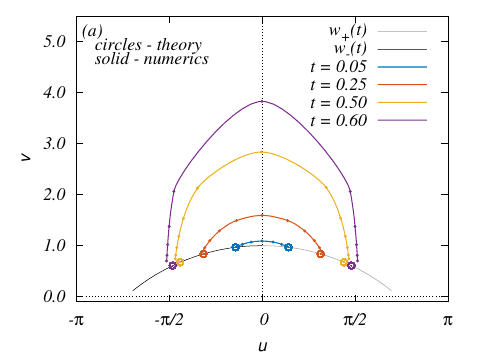}
\includegraphics[width=0.495\textwidth]{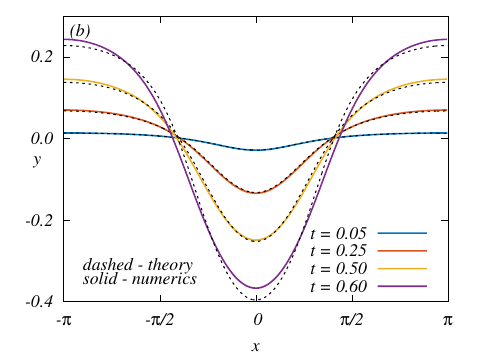}
\caption{(a) The location of the branch cut in the analytical continuation of the complex velocity $V$ \e{RVvar2} in the complex plane at several
instants of time. The initial conditions are given by~\e{Feqiniperiodic} and~\e{Rini0}
with $A=-\I$ and $a_0=\I$. The branch cuts recovered from the numerical simulations of the
equations~\e{Reqn4}--\e{Veqn4}. The filled circles show the positions of poles of rational
approximation, and the open circles correspond to the branch point locations given by the
analytic formula for $w_{\pm,i}(t)$ from Eq. \e{chi012periodic} at the respective time.
The solid black lines are the trajectories of $w_{\pm,i}(t)$ from the short cut approximation.
The difference in the position of the branch points estimated from the numerical simulation
and the short branch cut theory is $2.89\%$ at time $t=0.05$ ($\epsilon = 0.92$), and is
$6.45\%$ at time $t = 0.50$ ($\epsilon = 4.12$). The $w_{\pm,i}(t)$ from Eq. \e{chi012periodic}
give an excellent estimate for the branch points even for $\epsilon > 1$. (b) The spatial
profiles of the fluid surface from numerical simulation (solid lines), and short cut
approximation (dashed lines).
} \label{Aim1}
\end{figure}

\begin{figure}
\includegraphics[width=0.495\textwidth]{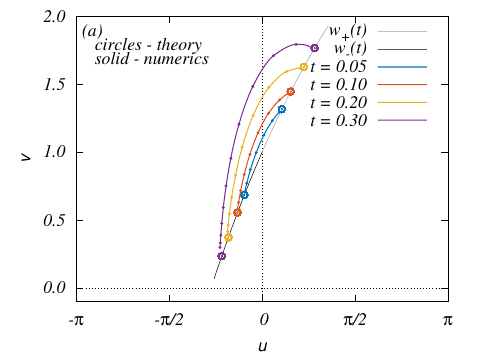}
\includegraphics[width=0.495\textwidth]{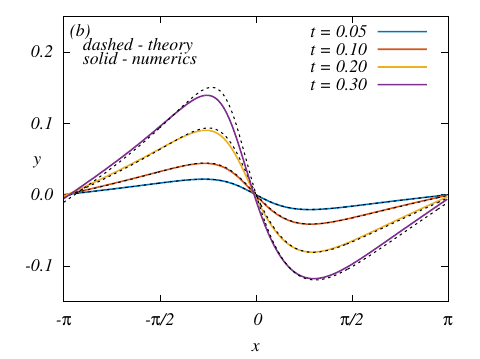}
\caption{(a) The location of the branch cut in the analytical continuation of the complex velocity $V$ \e{RVvar2} at different moments of time in the complex
plane for initial conditions given by~\e{Feqiniperiodic} and~\e{Rini0} with $A=1$ and
$a_0=\I$. The branch cuts recovered from full numerical simulations of Eqs.~\e{Reqn4}--\e{Veqn4}
is given by solid line. The filled circles represent the poles of the rational approximation of
the branch cut, and the open circles correspond to the branch point locations $w_{\pm,i}(t)$ from
the equation~\e{chi012periodic}. The gray line passing through $w = \I$ is the trajectory of
$w_{\pm,i}(t)$ as obtained from Eq.~\e{chi012periodic}. It is observed that for small time
$w_{\pm,i}(t)$ from Eq.~\e{chi012periodic} approximates the branch points to $3.31\%$
(relative error) at $t = 0.05$ with $\epsilon = 0.68$, and $7.69\%$ (relative error) at
$t = 0.25$ with $\epsilon = 1.18$. (b) The shape of free surface of the fluid at different
times: numerical simulation (solid lines), and the short branch cut approximation (dashed lines).
} \label{A1}
\end{figure}

These initial conditions result in a pair of branch points that move
according to  Eq.~\e{chi012periodic}. The direction of motion depends on
the argument of the complex constant $A$. In the simulations we chose
three values $A=1$, $A=\I$ and $A=-\I$. The initial pole of the complex
velocity, $V$, is located at $a_0=\I$. Generally $A$ can be the arbitrary
complex number and $a_0$ can be the arbitrary complex number from $\C^+.$
The Figures~\ref{Ai1}--\ref{A1} show the spatial profiles (right panel),
and the location of branch points (left panel) for both the branch cut
approximation of the Section~\ref{sec:narrowbranchcutperiodic}, and the
full numerical solutions. The branch points are located at $w=w_\pm(t)\equiv
w_{\pm,r}(t)+\I w_{\pm,i}(t)$, where $w_{\pm,r}(t)$ and $w_{\pm,i}(t)$ are
real--valued. At each time step the location is recovered from the
numerical simulations by the rational approximation procedure outlined in
Appendix \ref{sec:Rationalapproximation}. Additionally, $v_c=Im
(w_{-,r}(t))$ can be also determined from the asymptotic of the  exponential decay rate of the Fourier coefficients
 $   \hat z_k \sim e^{-v_c|k|}\quad\mbox{of $z(w)$ for}\quad |k|\to\infty,
$ 
~
see e.g. Ref. \cite{DyachenkoLushnikovKorotkevichPartIStudApplMath2016,LushnikovDyachenkoSilantyevProcRoySocA2017} for more details of that Fourier technique.
Eq.~\e{chi012periodic} provides the analytic formula for the location of
branch points in terms of the $\tau$ and $\chi$ in the branch cut
approximation. However, the dependencies of $\tau(t)$ and $\chi(w,t)$ are
given by an implicit relation that follows from the
Eqs.~\e{Feqiniperiodic}. For the sake of convenience we use the
approximate Eqs.~\e{chiw1} and~\e{taut1} which result in the explicit
expression~\e{chi012periodicw} for branch point locations in terms of $w$
and $t$.

We solve the implicit Eqs.~\e{FGsol} and~\e{chi0eq} for $\chi_0(w,\tau)$ at every
instant of time $\tau$ to determine the shape of the free surface $x(u,t)+\I y(u,t)$.
The extra conditions~\e{xpi} and~\e{Mass2} are used to find $c(\tau)$, and a subsequent
substitution in the Eqs.~\e{FGsol}--\e{FGsol2} and~\e{Feqiniperiodic} gives the shape
of the surface $z(u,t)$.

The summary of a comparison of the short branch cut approximation and the numerical solutions
is given below:

(a) For $A=\I$,  both branch points move along the imaginary axis as
follows from Eq.~\e{chi012periodic}. The lower branch point, $w_-= \I
w_{-,i}(t)$ moves downward from $w=a_0$, and the upper branch point $w_+=
\I w_{+,i}(t)$ is moves upward from $w=a_0$. Fig.~\ref{Ai1}a illustrates a
dependence of the vertical coordinate of $w_{-,i}(t)$ on time, as
determined from the Eq.~\e{chi012periodic} and the numerical simulations.
The positions of branch points are recovered from numerical simulations by
the procedure based on a least-squares rational approximation of complex
functions and is described in details in
Refs.~\cite{DyachenkoLushnikovKorotkevichPartIStudApplMath2016,
LushnikovDyachenkoSilantyevProcRoySocA2017,DyachenkoDyachenkoLushnikovZakharovJFM2019}.
The vertical coordinate of the lower branch point  is also estimated from the asymptotics of the
decay rate of the Fourier spectrum giving the same result. Fig.~\ref{Ai1}b shows the spatial
profiles of the free surface and a comparison of the short branch cut
approximation and full numerics.
It is seen that the spatial profile has a form of jet.
Also Fig.~\ref{Ai1}b shows the time dependence of the maximum error in the surface elevation $y(x,t)$ between the numerical solution and short branch cut approximation.

As discussed in all particular solutions of Sections  \ref{sec:narrowbranchcut} and \ref{sec:narrowbranchcutperiodic}, one of the branch points of the analytical solution in the short branch cut reaches the real line $w=Re(w)$ in a finite time meaning a formation of the singularity of the free surface in a finite time. This is exactly what is seen in Fig.~\ref{Ai1}a. However, we also see in   Fig.~\ref{Ai1}a that the full numerical solution seems does not produce a finite time singularity. Instead, the singularity appears to occurs at the infinite time $t\to \infty$. To quantify that statement we performed a fit to the stretcher exponential law  $v=a e^{-kt^{b}}$, where $a,b$ and $k$ are three real fitting constants. We find that $b=1.333\simeq 4/3$   provides the best fit as seen in Fig. \ref{fig:stretchedexponent}. Purely exponential fit $b=1$ is also shown providing not as good fit.  Another not as good fit is e.g. $b=2$, i.e. the Gaussian exponent (not shown in Fig. \e{fig:stretchedexponent}.
The detail discussion of the topic of finite time singularity is beyond the scope of this paper.

(b) For $A=-\I$, both branch points start to move in the horizontal direction, but unlike
the problem on infinite line $-\infty < x < \infty$, the branch points in periodic problem
develop vertical speed and approach the real axis. At later times branch cut
recovered from numerics is not short thus violating b the short
branch cut approximation. However, the positions of branch points recovered from short
branch cut approximation agree semi-quantitavely with
numerical simulations even at late times. 
Figure \ref{Aim1}b shows the spatial profiles
of the free surface at different times.

(c) For $A=1$,  both branch points start moving in the complex plane from the initial position at $w=\I a$ as illustrated in the
Figure~\ref{A1}a. Contrary to the other two cases, the positions of the branch points
are not symmetric with respect to the imaginary axis. The Figure~\ref{A1}b shows how the
shape of the free surface moves in time with increasing of steepness thus promoting overturning of the wave in a finite time.

We may conclude that the short branch cut approximation gives excellent results up to the
values of small parameter $\epsilon\gtrsim 0.9$, well--outside of the applicability region
for the short branch cut approximation~\e{narrowbranchcut1}.

\section{Conclusion and Discussion}
\label{sec:conclusion}
 The main result of this paper is the development of the short branch cut approximation both for the decaying BC \e{RVdecayingBC}  and the periodic BC \e{RVperiodicBC} for free  surface hydrodynamics. These equations in the moving complex frame are reduced to the fully integrable  complex Hopf equation \e{Hopfcompl} for the complex velocity $V(w,t)$ and the transport equation \e{ztauchigeneral} for the conformal map $z(w,t)$. These equations admit the infinite set of solutions easily constructed by the method of characteristics. Examples of such solutions are provided in Sections  \ref{sec:narrowbranchcut}  and  \ref{sec:narrowbranchcutperiodic}.  Section \ref{sec:numericalcomparison} demonstrated the excellent agreement between the analytical solutions of the short branch cut approximation  and the full numerical solution of Eqs.  \e{Reqn4}-\e{Veqn4} and \e{RVperiodicBC}.
 Examples of jets and overturning waves are shown in these solutions.

The results of Section \ref{sec:numericalcomparison}  appears quite striking because the analytical solutions of the short branch cut approximation agree with relatively good precision in several percents  with the solutions of Eqs.  \e{Reqn4}-\e{Veqn4}  even  when the parameter $\epsilon$
introduced in Eq.~\e{narrowbranchcut1} is not small while the derivation of Sections   \ref{sec:narrowbranchcut}  and  \ref{sec:narrowbranchcutperiodic} guarantees the applicability of the short branch approximation only for $\epsilon\ll 1$.  For future work we plan to analyze that efficiency of the short branch cut approximation for $\epsilon \gtrsim 1$ by addressing the corrections beyond that approximation outlined in Eqs. \e{VcRc1} and \e{UBnarrow}.

\conflict{We have no competing interests.}

\noindent {\textcolor{cyan}{Authors' contributions.}}   A.I.D. and V.E.Z.
contributed to developing analytical results. P.M.L.  and S.A.D.
contributed  to developing analytical results and numerical simulations.
All authors gave final approval for publication and agree to be held
accountable for the work performed therein.

\noindent {\color{blue} Data Accessibility.}
The codes used for producing Figures of Section \ref{sec:numericalcomparison} are publically available  at GitHub digital repository:  https://github.com/urrfinjuss/deep-water-full

\funding{ The work of P.M.L.  was supported by the National Science
Foundation, grant DMS-1814619. The work of V.E.Z. was supported by the
National Science Foundation, grant number DMS-1715323. The work of S.A.D.
was supported by the National Science Foundation, grant number
DMS-1716822. We thank support of Russian Ministry of Science and Higher
education Grant No. 075-15-2019-1893.
 Simulations were performed  at the Texas Advanced Computing
Center using the Extreme Science and Engineering Discovery Environment
(XSEDE), supported by NSF Grant ACI-1053575. P.M.L.  would like to thank
the Isaac Newton Institute for Mathematical Sciences, Cambridge, for
support and hospitality during the programme ``Complex analysis:
techniques, applications and computations" where work on this paper was
partially undertaken. }


\appendix

\section{Rational approximation for recovery of singularities}
\label{sec:Rationalapproximation}

In order to recover singularities of the functions $R$ and  $V$ in the complex
$w$-plane, we seek a rational approximation of the target function by
means of the Alpert--Greengard--Hagstr\"om (AGH) originally published
in Ref.~\cite{AGH2000}, and adapted for the water wave problem in Ref.~\cite{DyachenkoLushnikovKorotkevichPartIStudApplMath2016}.
We  outline the general approach to rational approximation of complex
functions, and refer the reader to the aforemetioned works for more
details.
See
also Ref.   \cite{DyachenkoDyachenkoLushnikovZakharovJFM2019} for the numerical demonstration of the high efficiency of that method.

AGH algorithm robustly recovers poles in
solution while branch cuts are approximated by a set of poles as
follows

\begin{equation}
\label{approx_cut} g(\zeta)=\frac{1}{2\pi} \int\limits_{C}
\dfrac{\rho(\zeta')\D\zeta'}{\zeta -\zeta'} \simeq  \sum\limits_{n =
1}^{N} \dfrac{\sigma_n}{\zeta -\zeta_n},
\end{equation}
where the function $g(\zeta)$ has s single branch cut along the
contour $C$ in the complex plane of $\zeta$ with $\rho(\zeta)$
being a jump of $g(\zeta)$ at the branch cut. R.h.s. of Eq.
\e{approx_cut} approximates $g(\zeta)$ by simple poles located at
$\zeta=\zeta_n\in C, \ n=1,\ldots, N$ with    the residues
$\sigma_n, \ n=1,\ldots, N$.

Given a $2\pi$--periodic function $f(w)$ on a real periodic
interval $w\in[-\pi,\pi]$, we may expand the periodic interval to the real
line, $-\infty < \zeta < \infty$, by a coordinate transformation
\begin{align}\label{zetadef}
\zeta = \tan\frac{w}{2}
\end{align}
which maps the stripe  $-\upi <Re(w)<\upi$ into  the complex $\zeta$
plane. Also $w\in\C^+(\C^-)$ imply that $\zeta\in\C^+(\C^-)$, see
also Ref.
\cite{DyachenkoLushnikovKorotkevichPartIStudApplMath2016} on more
details of the mapping \e{zetadef}.  $\zeta$ variable is convenient
to use in AGH algorithm
 (\cite{DyachenkoLushnikovKorotkevichPartIStudApplMath2016}) which is assumed below.

In the $\zeta$--variable, the function $g(\zeta) = f(w(\zeta)) - f(\pi)$ is
defined on the real line and decays to zero as $\xi\to\pm\infty$. The function
$g(\zeta)$ is suitable for rational approximation in the $\zeta$--variable,
and we seek two polynomials $P(\zeta)$ and $Q(\zeta)$ of degrees $N$ and $N+1$
respectively, such that:
\begin{align}
    \epsilon_N \equiv \int\limits_{-\infty}^{+\infty} \left|g(\zeta) - \frac{P(\zeta)}{Q(\zeta)}\right|^2\,d\zeta \to \min,
\end{align}
where minimization goes over the polynomial coefficients of $P(\zeta)$ and
$Q(\zeta)$.

After the optimal polynomials $P$ and $Q$ have been determined, the
resulting approximant gives accurate approximation to $g(\zeta)$ on the real
line. However, the ratio $P(\zeta)/Q(\zeta)$ defines a meromorphic function in
the complex $\zeta$--plane, and its singularities may be determined by
seeking the roots $\zeta_k$ of $Q(\zeta) = 0$. The residues at the poles are
given by $P(\zeta _k)/Q'(\zeta_k)$ and can be used to recover an approximation
to the Cauchy type integral as given by Eq. \e{approx_cut}.


\end{document}